\documentclass[preprintnumbers,showpacs,amsmath,amssymb,floatfix,prd,twocolumn,superscriptaddress,nofootinbib]{revtex4}
\usepackage{graphicx}
\usepackage{epsfig}
\usepackage{bm}
\usepackage{amsfonts}
\usepackage{epstopdf}
\usepackage{bm}
\usepackage{setspace}
\usepackage{neuralnetwork}

\begin{document}

\title{Artificial Neural Network Spectral Light Curve Template for Type Ia \\Supernovae and its Cosmological Constraints }

\author{Qiao-Bin Cheng}
\affiliation{Division of Mathematical and Theoretical Physics, \\ Shanghai Normal University,
	100 Guilin Road, Shanghai 200234, P.R.China}
\author{Chao-Jun Feng}
\email{fengcj@shnu.edu.cn} 
\affiliation{Division of Mathematical and Theoretical Physics, \\ Shanghai Normal University,
    100 Guilin Road, Shanghai 200234, P.R.China}

\author{Xiang-Hua Zhai}
\email{zhaixh@shnu.edu.cn} 
\affiliation{Division of Mathematical and Theoretical Physics, \\ Shanghai Normal University,
	100 Guilin Road, Shanghai 200234, P.R.China}

\author{Xin-Zhou Li}
\email{kychz@shnu.edu.cn} \affiliation{Division of Mathematical and Theoretical Physics,  \\ Shanghai Normal University,
    100 Guilin Road, Shanghai 200234, P.R.China}

\begin{abstract}
	The spectral energy distribution (SED) sequence for type Ia supernovae (SN Ia) is modeled by an artificial neural network. The SN Ia luminosity is characterized as a function of phase, wavelength, a color parameter and a decline rate parameter. After training and testing the neural network, the SED sequence could give both the spectrum with  wavelength range from 3000\AA~to 8000\AA~ and the light curve with phase from 20 days before to 50 days after the  maximum luminosity for the supernovae with different colors and decline rates. Therefore, we call this the Artificial Neural Network Spectral Light Curve Template (ANNSLCT) model. We retrain the Joint Light-curve Analysis (JLA) supernova sample by using the ANNSLCT model and obtain the parameters for each supernova to make a constraint on the cosmological  $\Lambda$CDM model. We find that the best fitting values of these parameters are almost the same as those from the JLA sample trained with the Spectral Adaptive Lightcurve Template 2 (SALT2) model. So we believe that the ANNSLCT model could be used to analyze a large number of SN Ia multi-color light curves measured in the current and future observational projects. 
\end{abstract}

\maketitle


\section{Introduction} \label{sec:intro}
Type Ia supernovae (SN Ia) have almost the same intrinsic brightness theoretically, so they could be used as a distance indicator on the cosmological scale, which is called the standard candle. But observations find that some SN Ia are a little brighter and some are a little dimmer. The difference between individual SN Ia magnitude could be described by some parameters, such as the color parameter, the shape parameter like the stretch or the decline rate \cite{Tripp:1998}\cite{Phillips:1993ng}. The explosion mechanism of SN Ia remains perplexing until today, so an SN Ia model needs to be constructed phenomenologically. Then, the parameters for each SN Ia could be obtained, for example, by fitting the light curve model to observational data.

In past decades, with exponentially increasing of the computing power of the hardware and the speedy development of the relevant algorithm, the artificial neural  network (ANN) becomes the most popular area of artificial intelligence field.  This technique has been used to solve  numerous practical problems in the world because ANN has excellent learning capacity. In this paper, by using the back-propagation ANN, we construct the spectral energy distribution (SED) sequence for SN Ia, which describes the evolution history of the SN Ia spectra. After training and testing the neural network, the SED sequence could give both the spectrum with  wavelength range from 3000\AA~to 8000\AA~ and the light curve with phase from 20 days before to 50 days after the  maximum luminosity for the supernovae with different colors and decline rates. Therefore, we call this the Artificial Neural Network Spectral Light Curve Template (ANNSLCT) model.

To train the SED sequence model, one should usually assume a functional form for the flux with  color and shape parameters,  as what is done in the Spectral Adaptive Lightcurve Template 2 (SALT2) model\cite{Guy:2007dv}. While, in the ANNSLCT model, there is no need to assume any relations. Details will be presented in the next section. 

In Ref.\cite{Cheng:2018nhz} , the authors have constructed the mean SED sequence with ANN by using the SN Ia spectrum data with and without the color parameter. In this paper, not only the additional light curve data will be included for constructing flux scale in various phase to improve the model, but also the shape parameter will be taken as another input variable for the ANN. The shape parameter that will be added is the decline rate  $\Delta m_{15}$, which  describes how fast the light curve declines in 15 days following B-band maximum luminosity.  High-z spectra is also used to expand the cover range of the wavelength low bound  from $3500$\AA  to $3000$\AA.

The supernova sample is also retrained by using the ANNSLCT model to obtain the parameters for each supernova. Then, we make a constraint on the cosmological  $\Lambda$CDM model. We find the best fitting values of these parameters are almost the same as those from the same sample but trained  by the SALT2 model. 

The structure of this paper is as follows. In Sec.\ref{sec:salt}, the SALT2 model will be briefly reviewed for later time comparison. In Sec.\ref{sec:ssm},  the SED sequence model and ANN will be described in detail.  In Sec.\ref{sec:data}, the data set including spectra and light-curves  for the training  will be described. In Sec.\ref{sec:training}, after combining the light curve and spectrum data, the neural network will be trained under different structures. In Sec.\ref{sec:results}, the training  results and the parameters of SN Ia will be presented. In Sec.\ref{sec:fit_cosmos}, the cosmological $\Lambda$CDM model will be constrained by using the ANNSLCT model. In Sec.\ref{sec:conclusion}, discussions and conclusions will be given.

\section{Brief review of SALT2}\label{sec:salt}

In the SALT2 model, the following functional form for the flux is used 
\begin{eqnarray}
F(p,\lambda) = x_0\bigg[\mathcal{M}_{0}(p,\lambda) + x_1\mathcal{M}_{1}(p,\lambda)\bigg] \exp[{c\times C_L(\lambda)}] \,,\label{equ:salt2}
\end{eqnarray}
where $x_0$ and $x_1$ are normalization and shape  parameters, respectively, and $x_1$ can be converted to the stretch or $\Delta m_{15}$ parameter, see Ref.\cite{Guy:2007dv}. Here $\lambda$ is the wavelength in the rest frame of SN Ia, and $p$ is the rest frame time before ($p<0$) or after ($p>0$) the date of maximum luminosity in the B-band, which is called the phase $p \equiv (t-t_{\text{max}}^B)/(1+z)$. $\mathcal{M}_{0}(p,\lambda)$ is the mean spectral sequence, and $\mathcal{M}_{1}$ is the first order deviation around the man sequence. $C_L(\lambda)$ is the mean color correction law, which is phase-independent and assumed to be third order polynomial law in the SALT2 model. 

To train the SALT2 model, one needs to minimize a $\chi^2$ function that measures the error between the model of Eq.(\ref{equ:salt2}) and the photometric and spectroscopic data sample. The SED sequence in the SALT model will be treated as $\mathcal{M}_0$ for the stretch $s=1$, while the difference between the SED sequence of an SN Ia with stretch $s=1.1$ and $\mathcal{M}_0$ is treated as $\mathcal{M}_1$. Thus, the SALT2 model is a linearized version of the SALT model. The SALT2 model needs more than 3000 parameters to fit with Gauss-Newton procedure, see Refs.\cite{Guy:2007dv}\cite{Mosher:2014gyd} for details. The trained SED sequence model covers the phase range of $[-20,50]$ days and a spectral range of $[2000,9200]$\AA~with resolution of $60$\AA~for  $\mathcal{M}_0$.  For $\mathcal{M}_1$, a low resolution is used\cite{Guy:2007dv}.

\section{SED sequence with ANN}\label{sec:ssm}

\begin{figure}[thp]
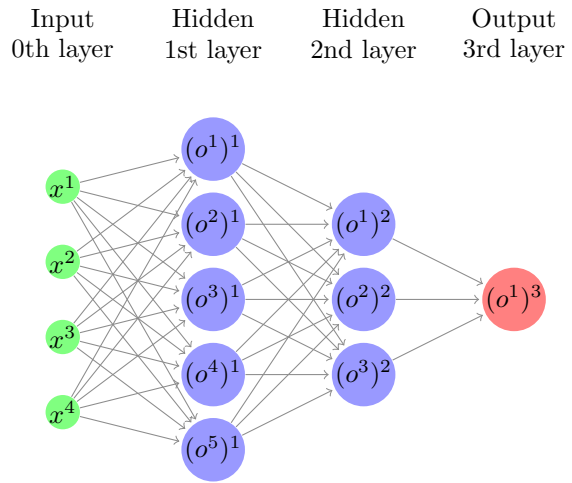

	\begin{center}
		\begin{neuralnetwork}[nodespacing=10mm, layerspacing=20mm,
			maintitleheight=2.5em, layertitleheight=5.em,
			height=4.5, toprow=false, nodesize=10pt, style={},
			title={}, titlestyle={}]
			\newcommand{\nodetextclear}[2]{}
			\newcommand{\nodetextx}[2]{$x^#2$}
			\newcommand{\nodetexty}[2]{$y_#2$}
			\newcommand{\nodetexto}[2]{$(o^#2)^#1$}
			\inputlayer[count=4, bias=false, title=Input\\0th layer , text=\nodetextx]
			\hiddenlayer[count=5, bias=false, title=Hidden\\1st layer, text=\nodetexto] \linklayers
			\hiddenlayer[count=3, bias=false, title=Hidden\\2nd layer, text=\nodetexto] \linklayers
			\outputlayer[count=1, title=Output\\3rd layer, text=\nodetexto] \linklayers
		\end{neuralnetwork}
		\caption{Typical structure of an ANN.}
		\label{fig:net}
	\end{center}
\end{figure}

The ANN that will be constructed is called the back-propagation neural network, which has been used in astronomy and physics.  For example, by using ANN, the type of a supernova could be classified into e.g.Ia,  Ib,  II, etc., see \cite{Graff:2013cla} and references therein. The structure (or topology) of an ANN could be described as Fig.\ref{fig:net}.
One can see that the ANN has four layers. The first layer is called the input layer or the input for short, while the last layer is called the output layer or the output for short. The layers between the input and output are all hidden layers, and there are two hidden layers in Fig.\ref{fig:net}. There are a number of neurons at each layer. For example, there are five and three neurons at the first and second hidden layers respectively. The number of neurons at the input and, output are called the dimensions of the input and the output. 

The SED sequence will be modeled by an ANN with 4-dimensional input , i.e. $(x^1,x^2,x^3,x^4) = (p,\lambda,c,\Delta m_{15})$ and 1-dimensional output as the SED sequence flux. The ANN network seems like a nonlinear function with four independent variable, i.e.  $F^{\text{ANN}}(p,\lambda,c,\Delta m_{15})$.  
   
In Fig.\ref{fig:net}, each neuron in one layer of the ANN has directed connections to the neurons of the subsequent layer. So, this kind of ANN is also called the completely-fully connected neural network. The weights on these connections could be regarded as the parameters of the ANN, and their values will be obtained by training the network. Here training  means making the output almost the same as the observational data.   

On each neuron, there is also an activation function, which is often taken as a nonlinear function like a sigmoid function, a tangent hyperbolic function etc.. Just like that in the biological neural network, the activation function decides the output of a neuron, whether a neuron will be activated or not. And the argument of the activation function is the weighted sum of neuron outputs at the previous layer and further with a bias added.

 Outputs of all neurons at a layer will be sent to the subsequent layer. Due to the non-linearity of the activation function, the ANN could be able to describe highly nonlinear function like the SED sequence. See Ref.\cite{Cheng:2018nhz} for details of the ANN constructing structures.

The aim of training an ANN is to minimize a cost  function with training samples. The cost function describes the error between the output and the samples. For the spectra samples, the cost function is given by $E = \mathbf{ e^TC^{-1} e}/2$,  with $\mathbf{C}$  the covariance matrix of the observational flux and $\mathbf{e}  = F^{\text{obs}} - F^{\text{ANN}}$. For the light curve samples, there is a little different cost function, which will be described in the next section. The total cost function will be the sum of all cost functions for each data sample.
To find the minimum of the cost function, we will take the Levenberg-Marquardt (LM) algorithm, which converges to the minimum much faster than the steepest gradient descent algorithm and has much less calculations than the quasi-Newton methods. In each step $s$, the weight and bias $W$ will be updated as 
\begin{equation}\label{eq:updateWeights}
W_{s+1}=W_s-\mathbf{\left(J^TC^{-1}J + \mu I \right)^{-1}}\mathbf{J^TC^{-1}e}\,,
\end{equation}
 where $\mathbf{I}$ is the identity matrix, $\mu$ is the combination coefficient that could be changed adaptively during the training procedure. The Jacobi matrix $\mathbf{J} \in \mathcal{R}^{M\times N}$ is defined as
\begin{equation}
\begin{pmatrix} 
\frac{\partial e_1}{\partial W_1} & \frac{\partial e_1}{\partial W_2} &\cdots& \frac{\partial e_1}{\partial W_N}\\ 
\frac{\partial e_2}{\partial W_1} & \frac{\partial e_2}{\partial W_2} &\cdots& \frac{\partial e_2}{\partial W_N}\\ 
\cdots&\cdots&\cdots&\cdots\\
\frac{\partial e_M}{\partial W_1} & \frac{\partial e_M}{\partial W_2} &\cdots& \frac{\partial e_M}{\partial W_N}\\ 
\end{pmatrix}\,,
\end{equation}
where $N = \sum_{l=1}^{L+1} N^{l}(N^{l-1}+1)$ is the total number of  weights ( including bias), see Ref.\cite{Cheng:2018nhz} for more details.

\section{Data Description} \label{sec:data}
\subsection{low-z spectra}

This low-z data set contains 1787 spectra of 238 SN Ia(about 4600 thousand data points)\citep{Blondin:2012ha,Matheson:2008pa,Branch:2003hk,Jha:1999sm,Krisciunas:2011sn,Li:2003wja,Foley:2009wk,Hicken:2007ap}.  In Ref.\cite{Cheng:2018nhz} the authors have analyzed the distribution properties of this data set, in which most data samples have high signal-to-noise ratio. So the low-z spectra data are the main data to train  the ANNSLCT model.

\subsection{high-z spectra}
The ESO/VLT 3rd year Type Ia supernova data set\cite{Balland:2009ka} contains 139 spectra of 124 SN Ia. It constitutes the high-z spectra of training data. The redshift ranges from z = 0.149 to z = 1.031 in this set. Two outlier spectra for SN 05D4ay are dropped.  The low-z set has poor ability to cover in spectra in the UV band, while the high-z spectra have better performance in this band. However, this high-z data set has low quality, so the blue end wavelength is expanded only to $3000$\AA.

\subsection{light curves}
The light curve data we used is the same as those in the JLA  sample\cite{Betoule:2014frx}, which contains 740 SN Ia. This set includes 130 nearby SN Ia\cite{Conley:2011ku}, the recalibrated SDSS-II light-curves of 368 SN Ia\cite{Betoule:2014frx}, 239 SN Ia of SNLS  and 9 very high redshift(0.7$<$z$<$1.4) SN Ia observed by HST. The light curve data set is used to restrict the SED sequence scale for individual supernova and constrain the cosmological model for the ANNSLCT model.

\section{Training process} \label{sec:training}

The luminosity of SN Ia could be related to two parameters\cite{Tripp:1998}\cite{Riess:1996pa}\citep{Hamuy:1996ss}. One of them is the decline rate $\Delta m_{15}$, which measures the descending rate of the SN Ia luminosity 15 days after it reaches the maximum in the B-band. The other is the color parameter $c$, which is defined as the difference of the maximum magnitudes between the B-band and the V-band, and it also measures the supernova temperature as its maximum luminosity. The relation between the distance modulus $\mu$ and these parameters is usually assumed  as the following empirical formula:
\begin{equation}\label{eq:mu}
	\mu = m_B - (M_B +\alpha \times \Delta m_{15} + \beta \times c)\,,
\end{equation}
where $m_B$ is the observed maximum magnitude in rest-frame B-band, $\alpha$ and $\beta$ are two constants,  and the absolute magnitude $M_B$ is some function of the host galaxy mass\citep{Sullivan:2011kv}\cite{Johansson:2012si}, which could be approximated by the following piecewise function\cite{Conley:2011ku}:
\begin{equation}
	M_B = \left\{ \begin{array}{ll}
			M_B^{\ast} \, & \textrm{if } M_{host} < 10^{10} M_{\odot}\,, \\
			M_B^\ast + \Delta_M \,& \textrm{if } M_{host} \geqslant 10^{10} M_{\odot}\,,
	\end{array}\right.
\end{equation}
after the host stellar effect is corrected. In contrast, the hypothesis quantified by a linear model in the JLA samples is given by\cite{Betoule:2014frx}
\begin{equation}\label{eq:mu}
\mu = m_B - (M_B -\alpha' \times X_1 + \beta \times c)\,,
\end{equation}
where $X_1$ is the stretch factor, see Ref.\cite{Betoule:2014frx}. 

As we introduced in Sect.\ref{sec:ssm}, the SED sequence is modeled by an ANN with phase, wavelength, color and $\Delta m_{15}$ as its input. The calculation of $\Delta m_{15}$ is easy and the role it plays is almost equivalent to the stretch factor $X_1$. The information of SN Ia color, $\Delta m_{15}$ and the maximum date could not be obtained without an SN Ia light curve template model, which is described by the network and needs to be trained. So, at the beginning of training, these parameters  are estimated for each individual supernova and then they are adjusted and corrected in each iteration during the training process. In this way, one doesn't need to assume a relation between these parameters and the flux as that used in the SALT2 model, see Eq.(\ref{equ:salt2}). All of the relations are encoded in the ANN model.

The treatment of the light curve data is a little different from that of spectral data, because the flux of the light observed in the light-curve will be through a filter. And its value  $\phi(p,c,\Delta m_{15})$ could be obtained by the sum of SED sequence:
\begin{equation}\label{eq:fluxExpression}
\phi(p,c,\Delta m_{15}) \propto  \sum_{\lambda _{i}} F(p,\lambda_i,c,\Delta m_{15})T(\lambda_i)\lambda_i\Delta\lambda\,,
\end{equation}
where $T(\lambda)$ is the overall instrument response function for some band, for example $T_b(\lambda)$ is for the B-band. Therefore, the error function  for training could be defined as  $e = \phi^{obs}  - \phi^{model}$ . 

Data will be split into two parts, 80\% of which are for the training set, and the rest are for the test set in order to avoid over-fitting. To optimize the model, the so-called the second-order algorithm\cite{Minsky}, i.e. the LM algorithm, is realized. And we also use the Graphics Processing Unit (GPU) to accelerate the training process.

\section{Results} \label{sec:results}

The structures of the neural network we used are summarized  in the first column of Table.\ref{tab:trainingResults}. The root mean squared error(RMSE) and the mean absolute error(MAE) are given in the third and the forth columns. The second column of Table.\ref{tab:trainingResults} is the number of parameters or weights of the network.
\begin{table}[ht]
	\begin{tabular}{rccc}
		\hline
		\hline
		Structure & Number & Training set & Test set \\
		4-20-20-20-20-1 & 1381 & 0.3963/0.2091 & 0.4156/0.2194 \\
		4-19-19-19-19-1 & 1255 & 0.4062/0.2210 & 0.4191/0.2257 \\
		4-18-18-18-18-1 & 1135 & 0.4100/0.2204 & 0.4474/0.2338 \\
		4-17-17-17-17-1 & 1021 & 0.4257/0.2270 & 0.4403/0.2356 \\
		4-15-15-15-15-1 &  811 & 0.4336/0.2295 & 0.4511/0.2412 \\
		\hline
		\hline
	\end{tabular}
	\caption{The RMSE/MAEs over the training and test sets for various nets(The second column is number of weights).}\label{tab:trainingResults}
\end{table}
From Table.\ref{tab:trainingResults}, one can see that only about 1000 parameters are needed for a network with simple four hidden layers to describe so complicated template model. We have constructed more complicated structures for the network and we find the improvements are very limited.

\subsection{spectra}
\begin{figure}
		\begin{center}
			\includegraphics[height=0.15\textheight,width=0.5\textwidth,angle=0]{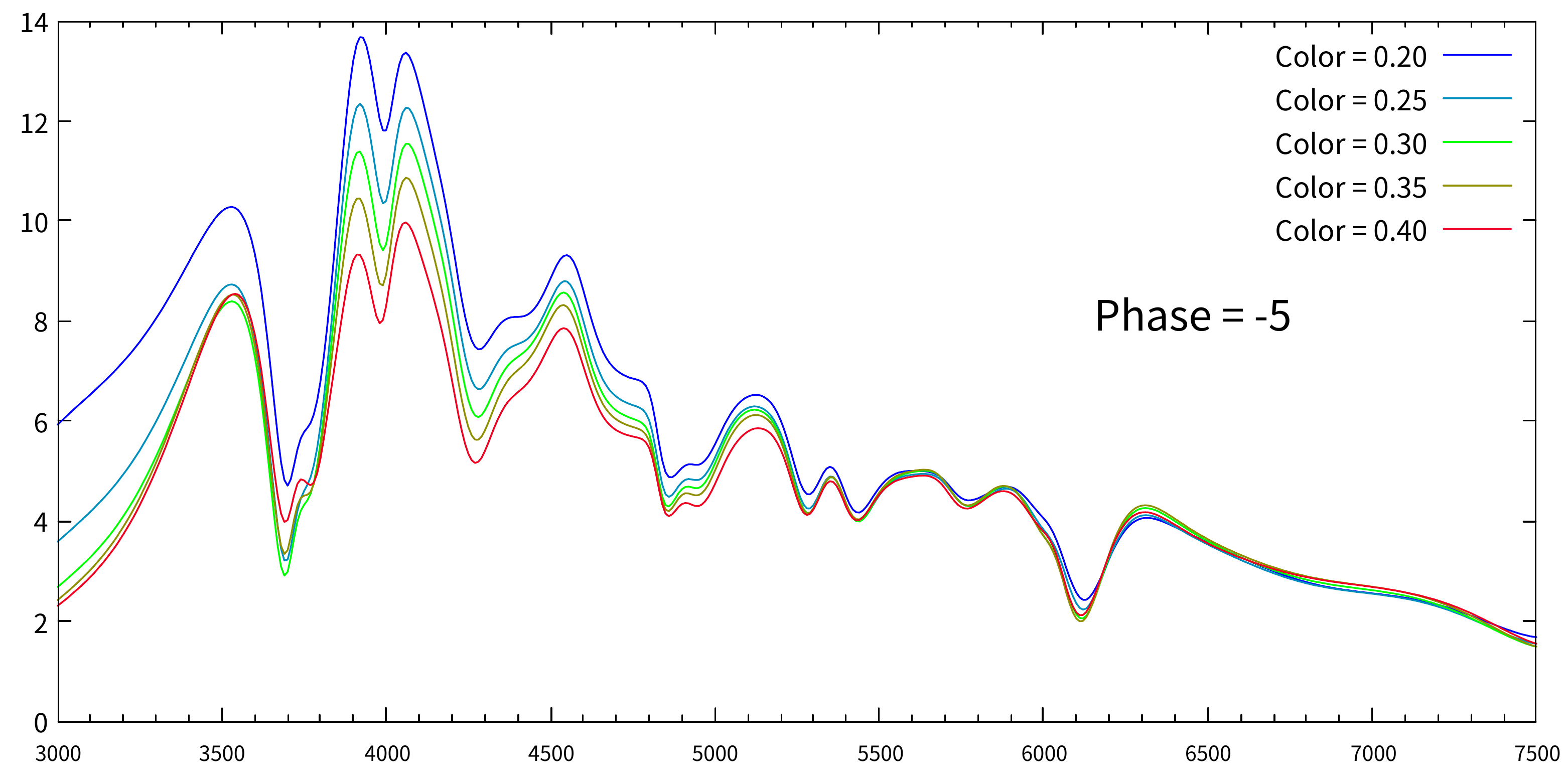}
			\includegraphics[height=0.15\textheight,width=0.5\textwidth,angle=0]{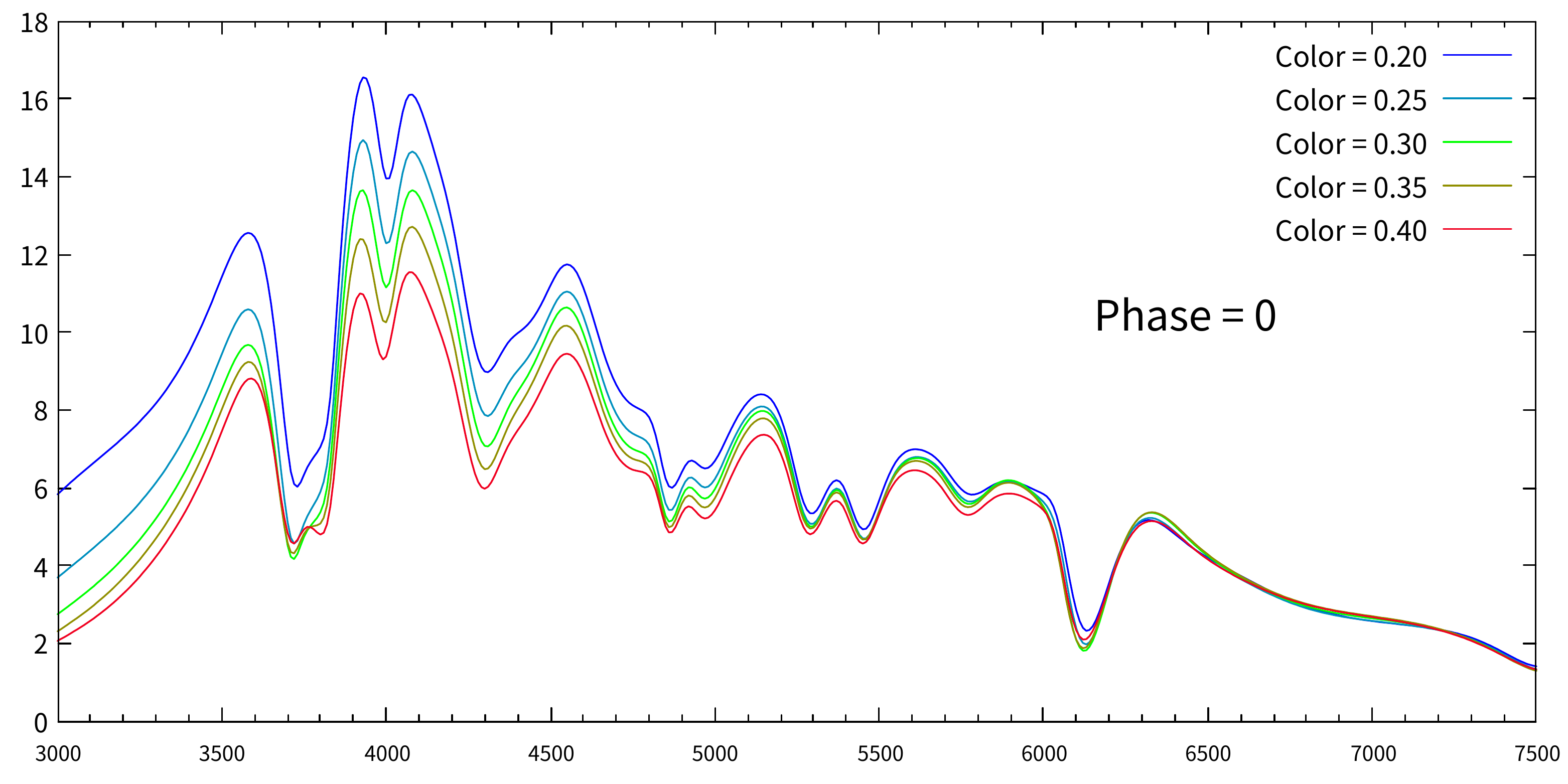}
			\includegraphics[height=0.15\textheight,width=0.5\textwidth,angle=0]{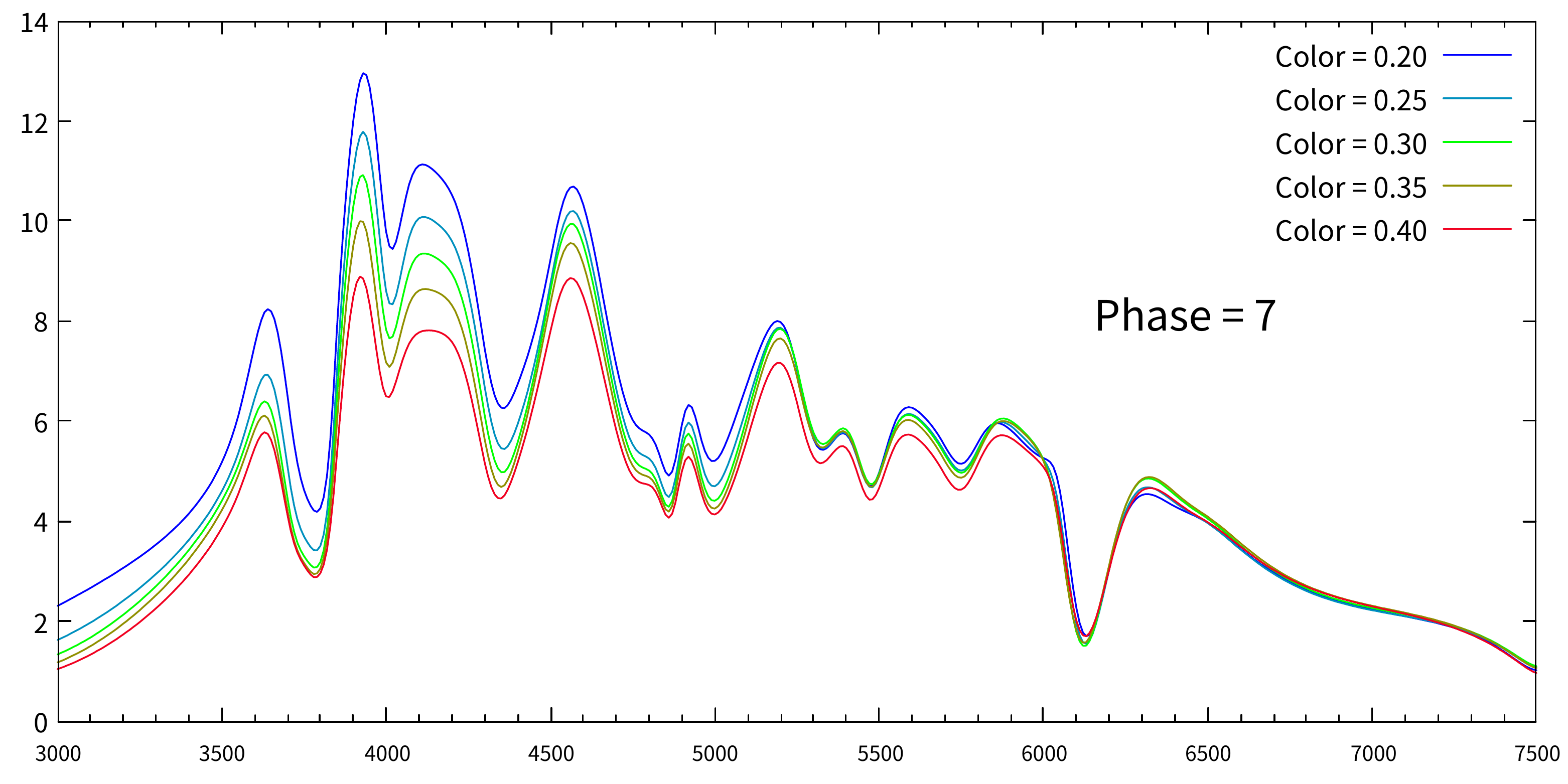}
			\includegraphics[height=0.15\textheight,width=0.5\textwidth,angle=0]{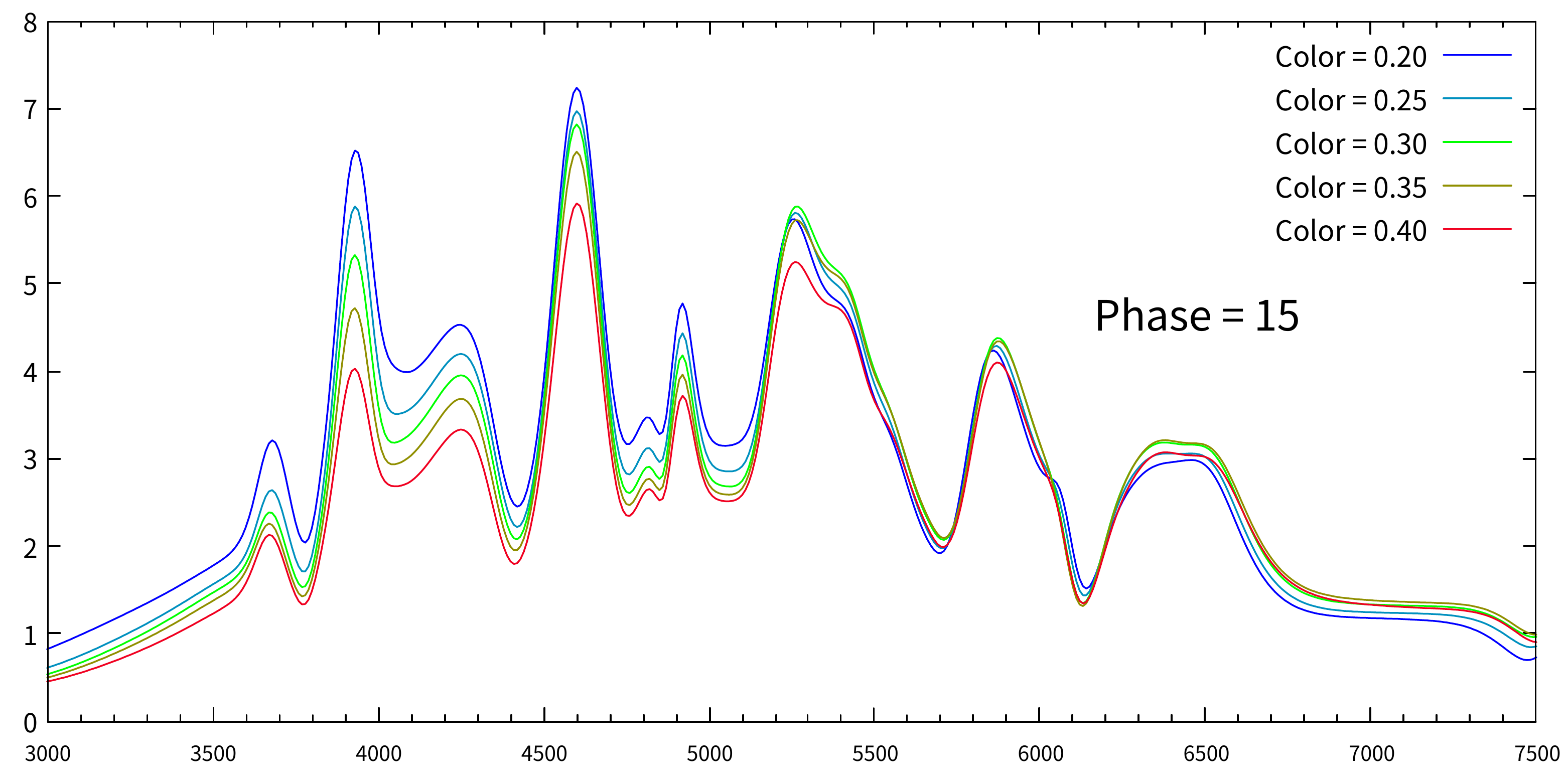}
		\caption{Comparison of several color spectra(these spectra have the same $\Delta m_{15}$).}
			\label{fig:compareColor}
		\end{center}
	\end{figure}

\begin{figure}
		\begin{center}
			\includegraphics[height=0.15\textheight,width=0.5\textwidth,angle=0]{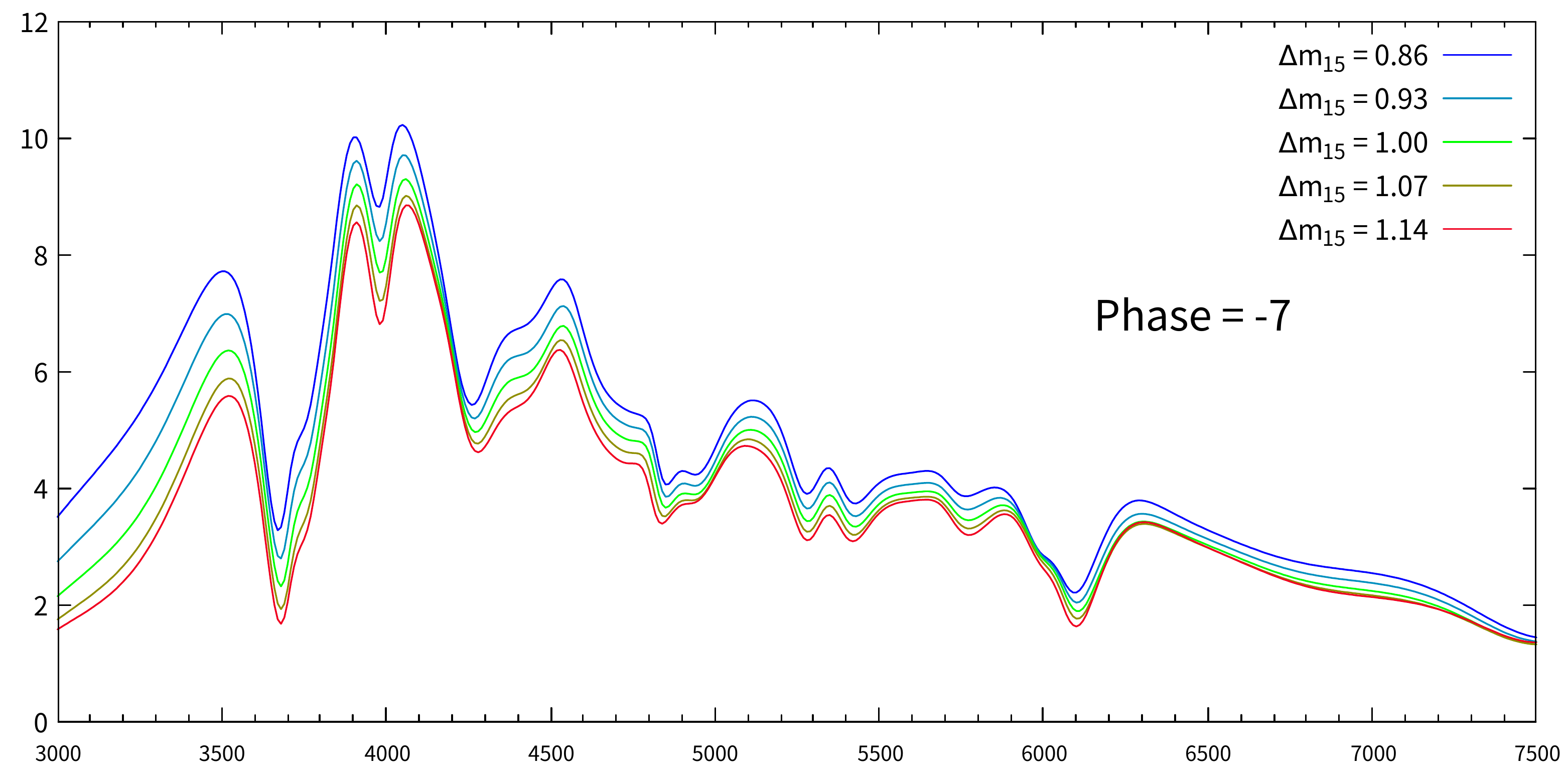}
			\includegraphics[height=0.15\textheight,width=0.5\textwidth,angle=0]{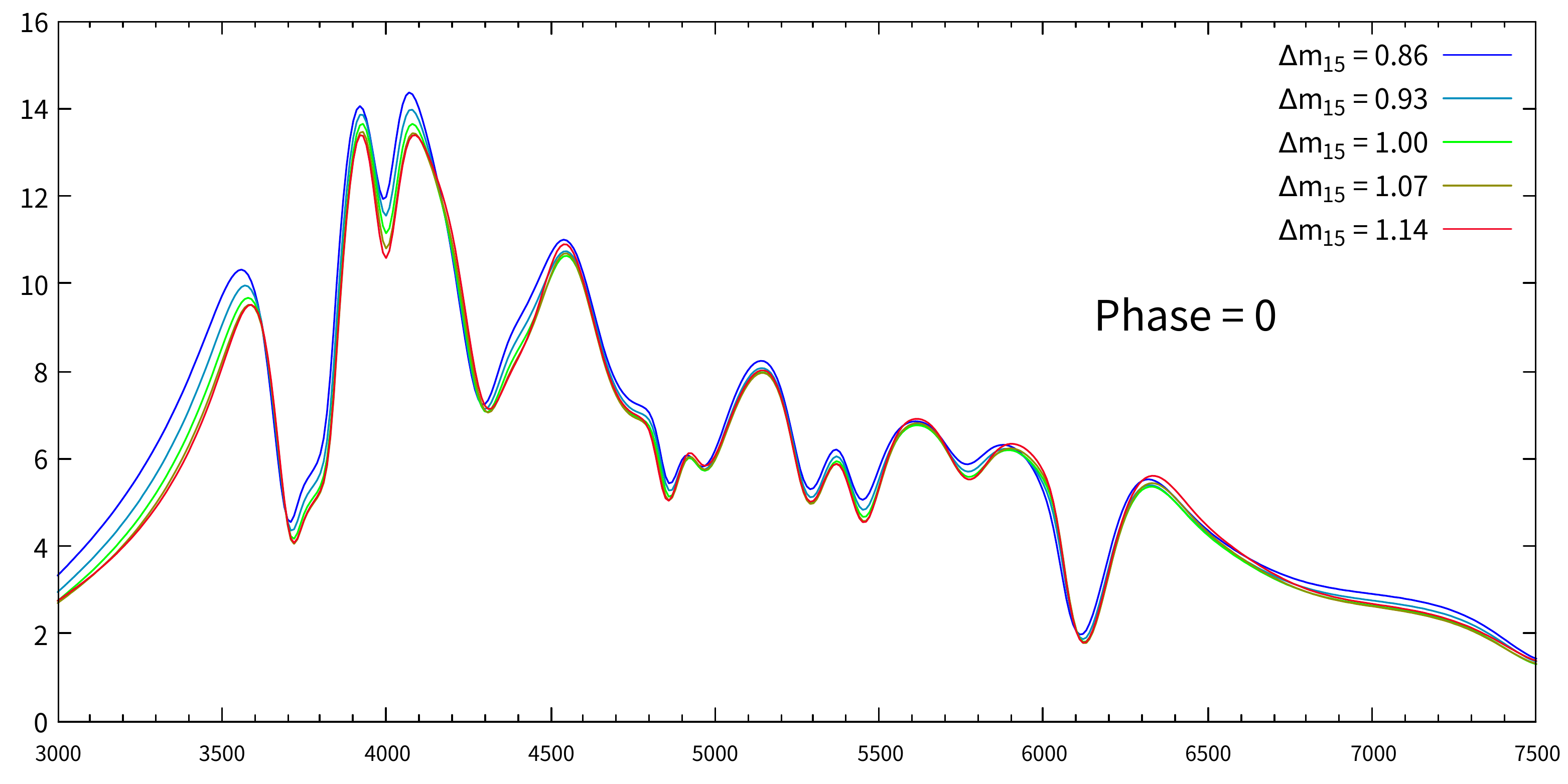}
			\includegraphics[height=0.15\textheight,width=0.5\textwidth,angle=0]{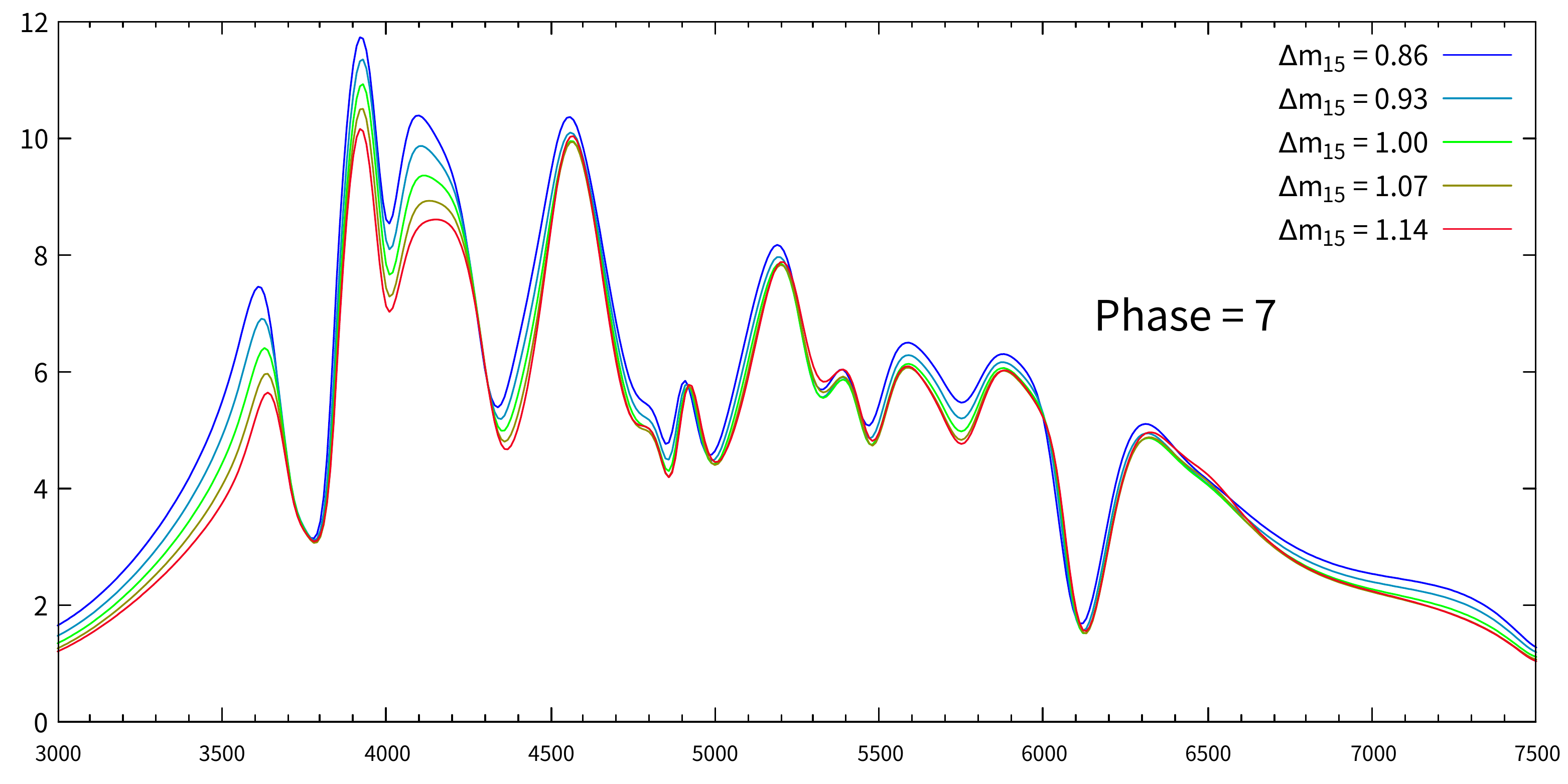}
			\includegraphics[height=0.15\textheight,width=0.5\textwidth,angle=0]{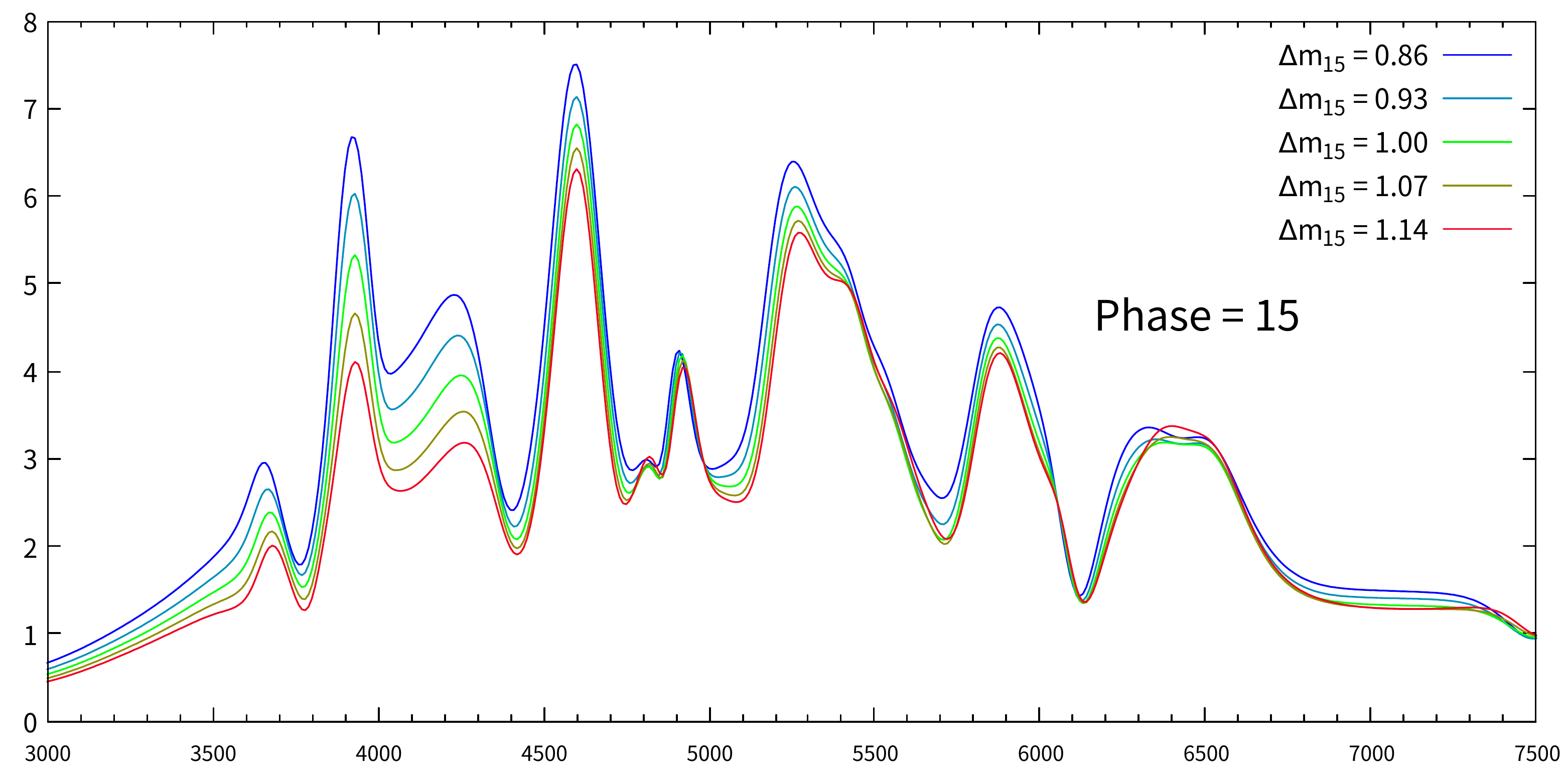}
		\caption{Comparison of various $\Delta m_{15}$ spectra(These spectra have the same $c$).}
			\label{fig:compareDm15}
		\end{center}
	\end{figure}

In Fig.\ref{fig:compareColor}, the evolution of spectra for different colors with the same  $\Delta m_{15}$ is plotted. One can see that the spectra are almost independent of the color in  the red part of spectrum, while they are depressed by a large color value in the blue part. 
In Fig.\ref{fig:compareDm15}, the evolution of spectra for different $\Delta m_{15}$ with the same color is plotted. It shows that after explosion of the supernova, the differences between the spectra become more and more obvious. This is consistent with the definition of  $\Delta m_{15}$. And from  Fig.\ref{fig:compareColor}, one can also see that there is more differences between the spectra when the phase is 15 days after its maximum. This suggests a faster decline of magnitude due to the quick falling of the SN's temperature.

\subsection{light curve}
\begin{figure}
		\begin{center}
			\includegraphics[height=0.20\textheight,width=0.5\textwidth,angle=0]{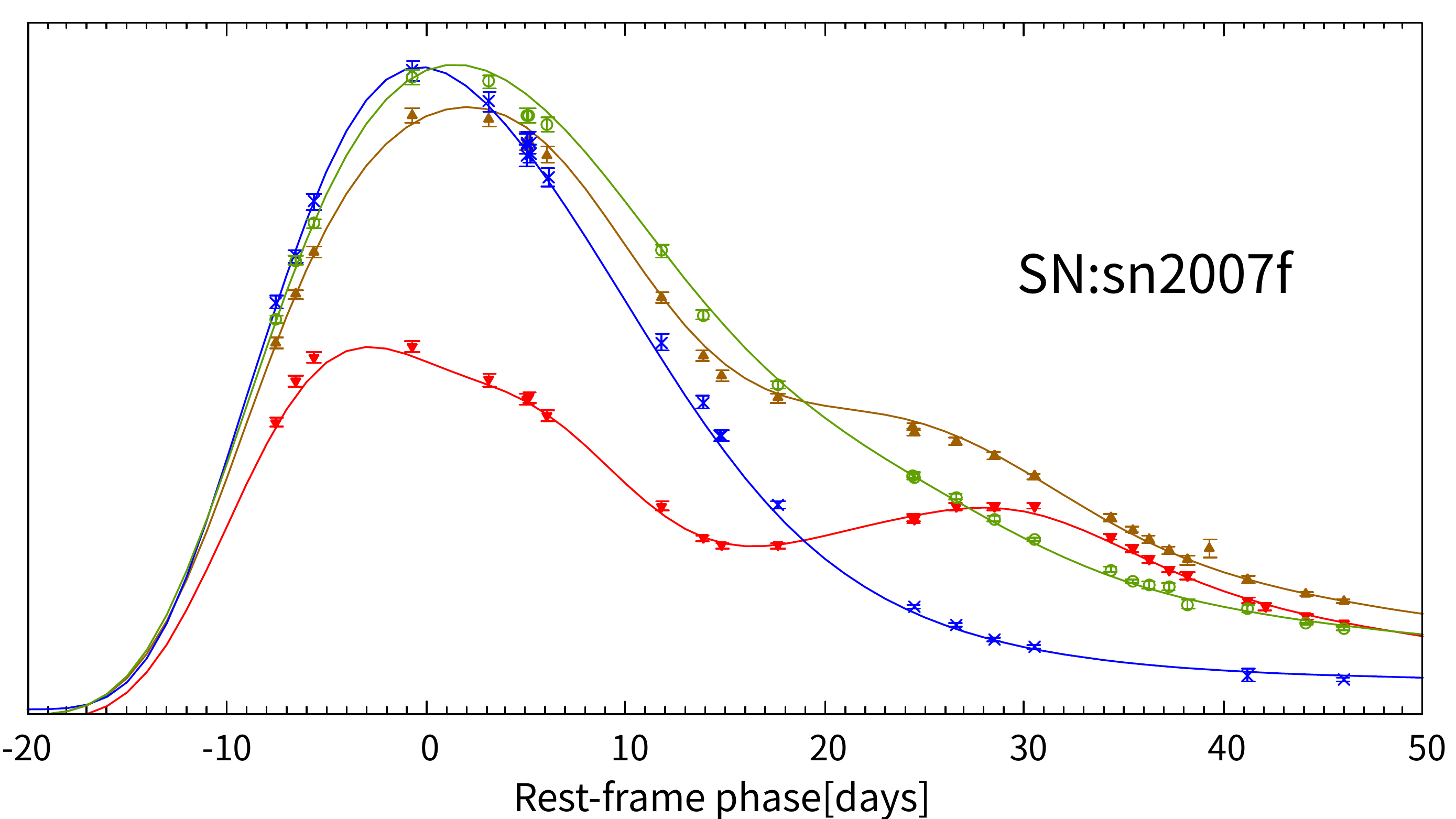}
			\includegraphics[height=0.20\textheight,width=0.5\textwidth,angle=0]{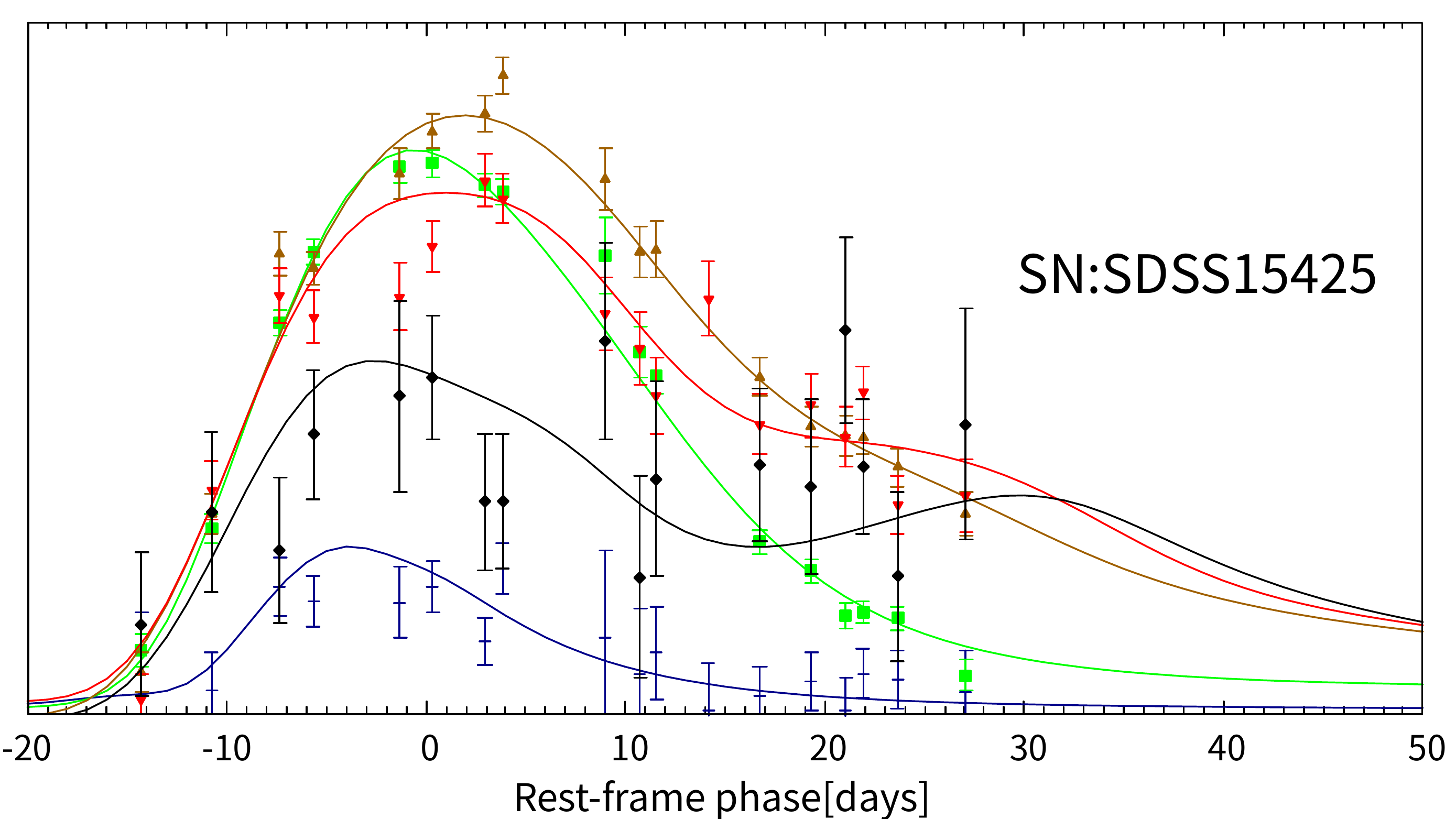}
			\includegraphics[height=0.20\textheight,width=0.5\textwidth,angle=0]{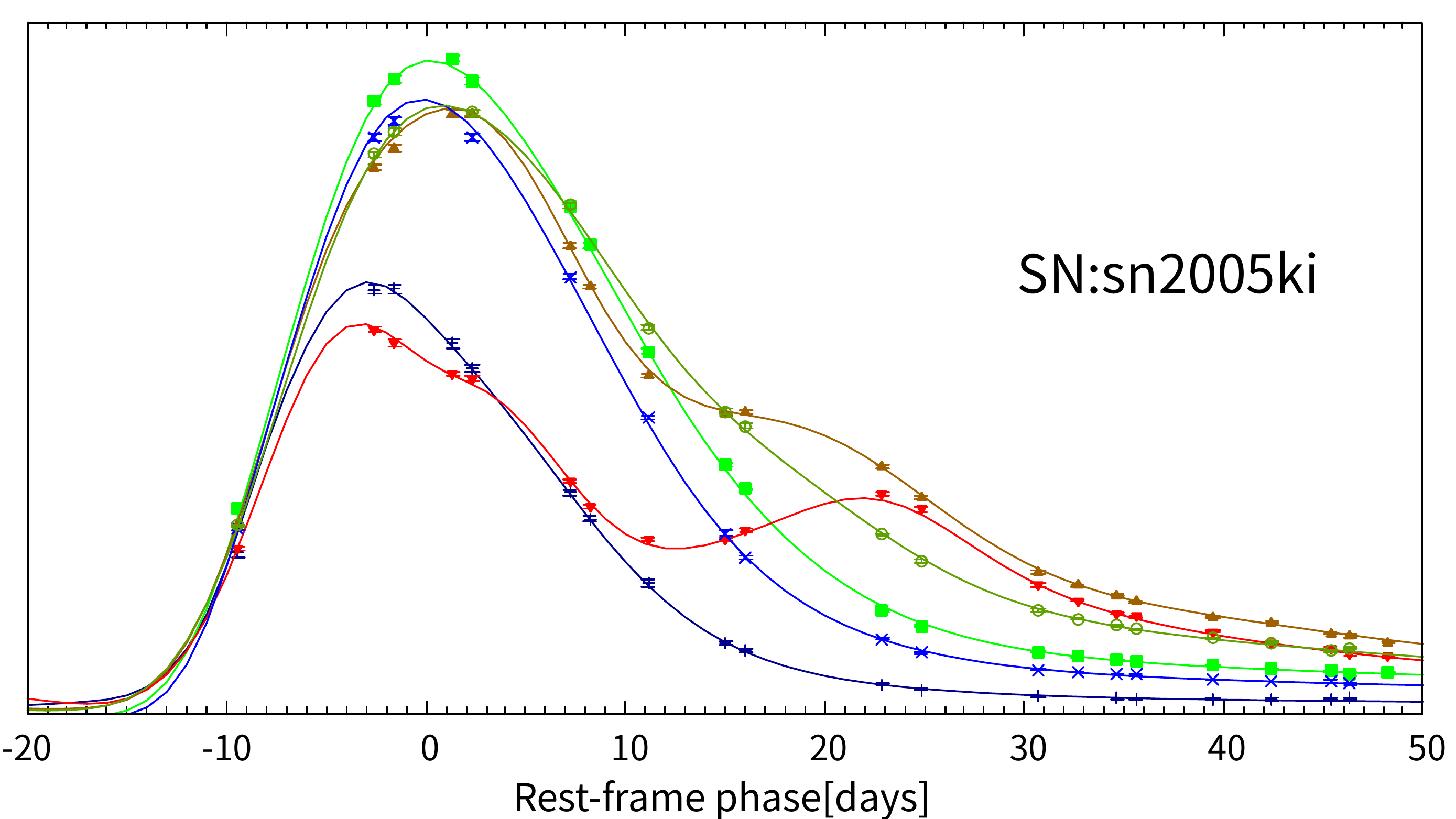}
			\caption{Light-curves of some supernovae. The dots denote the observed data, while the curves are generated by the template model.}
			\label{fig:lightcurveAndObserveData}
		\end{center}
	\end{figure}

As we mentioned before,  when the light-curve data is also included in the training process, one needs to estimate the parameters of each individual supernova and then update them in each iteration. And the flux is obtained by summing the energy distribution template model with the response function as its weights, see Eq.(\ref{eq:fluxExpression}).  In Fig.\ref{fig:lightcurveAndObserveData}, light-curves of several supernovae from the template model and their observed values are plotted. 
For the supernova with high quality data (e.g., SN sn2007f), the light-curves generated by the model agree very well with the observational values. While for the supernova with  low signal-to-noise ratio data (e.g., SN SDSS15425), the net can also draw their light-curves.

\subsection{parameters}

\begin{figure}
		\begin{center}
			\includegraphics[height=0.3\textheight,width=0.5\textwidth,angle=0]{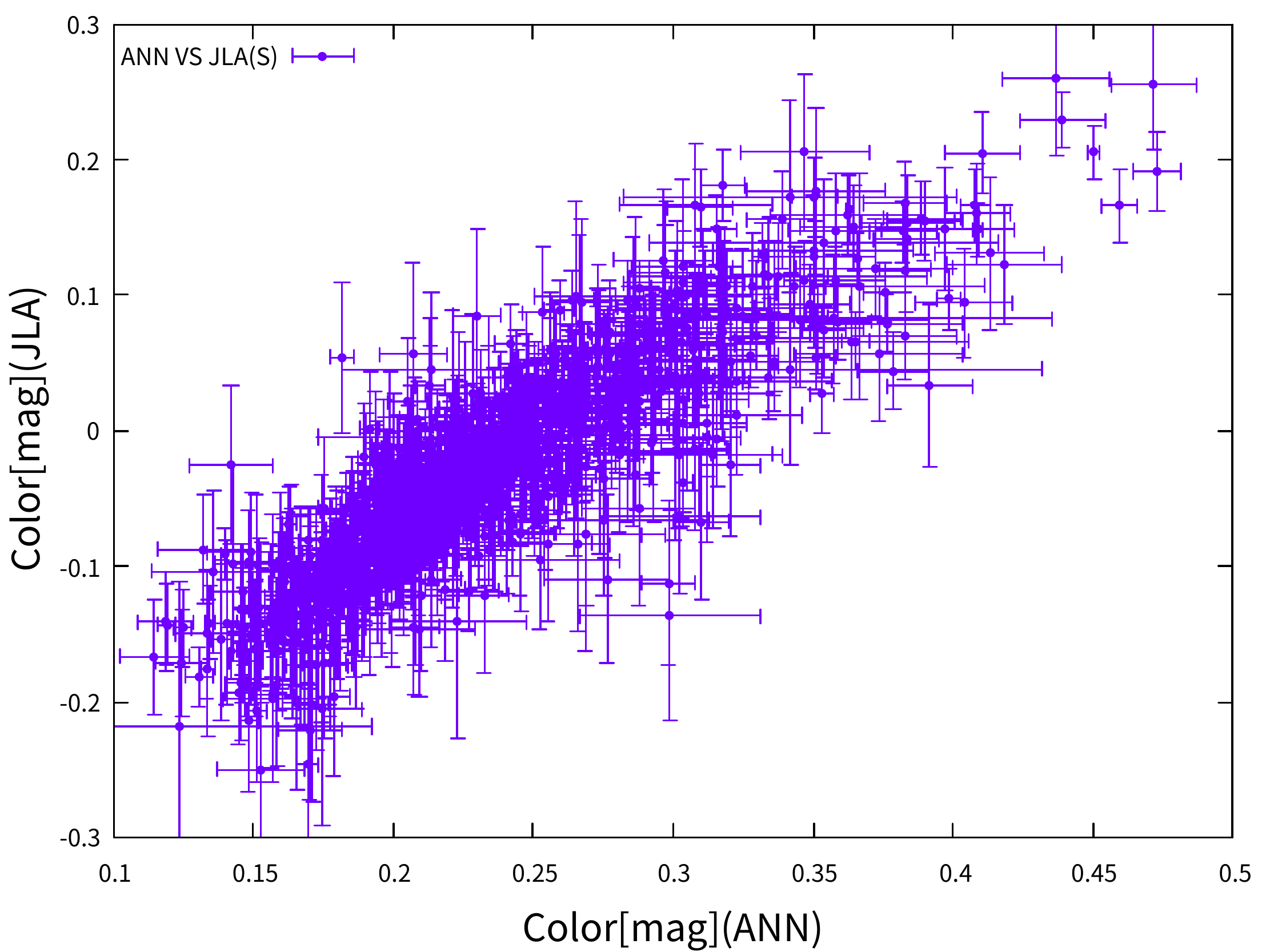}
			\caption{Comparison of the color parameter.}
			\label{fig:Color_ANN_VS_JLA}
		\end{center}
	\end{figure}

\begin{figure}
		\begin{center}
			\includegraphics[height=0.3\textheight,width=0.5\textwidth,angle=0]{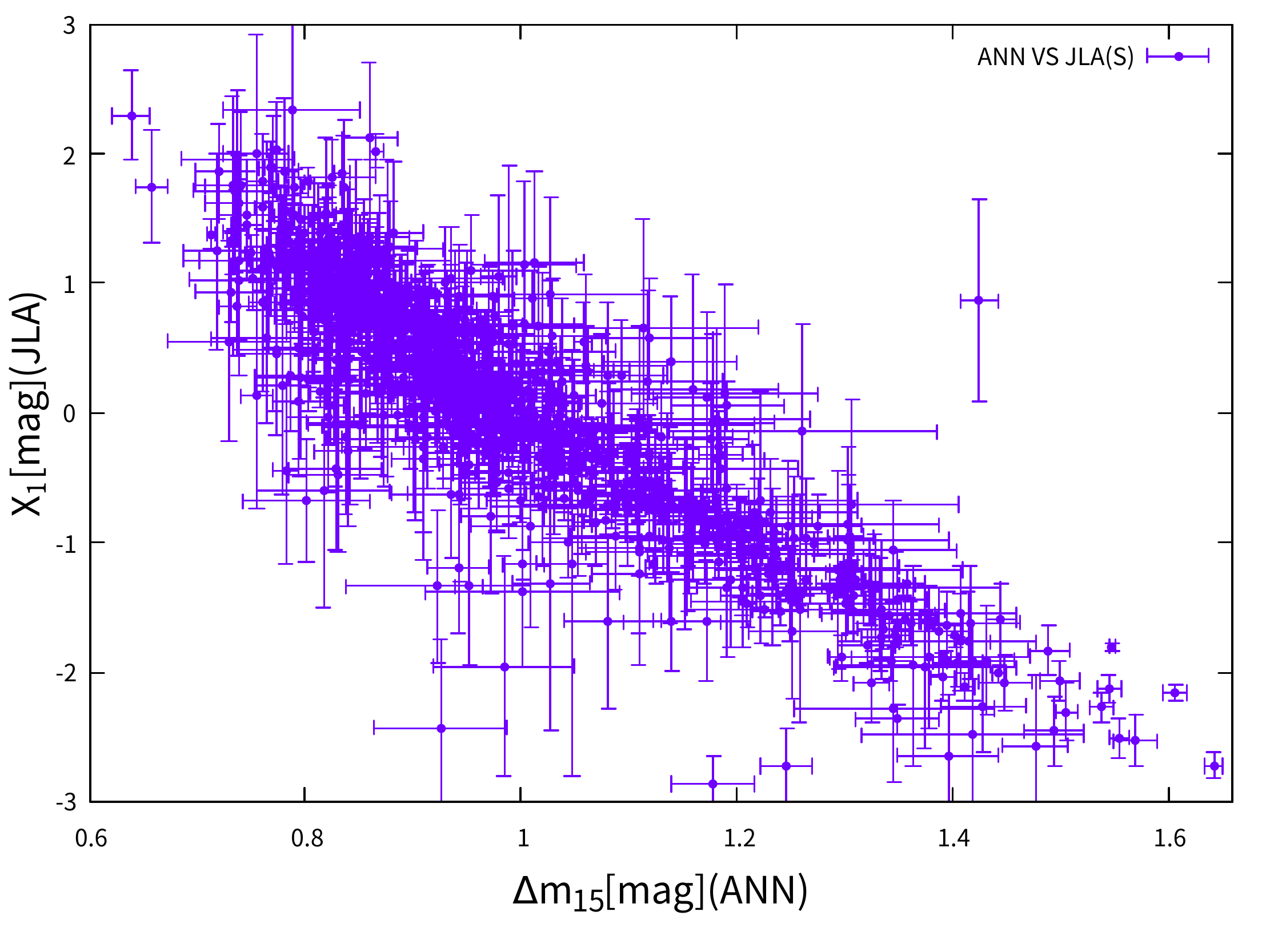}
			\caption{Comparison of $\Delta m_{15}$ and the stretch parameter.}
			\label{fig:Dm15_ANN_VS_JLA}
		\end{center}
	\end{figure}

	The parameters of SN Ia are crucial to determine their absolute magnitude, so it is necessary to compare our results with those from the other methods like SALT2. In Fig.\ref{fig:Color_ANN_VS_JLA}, the parameters $c$ constrained by the ANNSLCT model and those by the SALT2 (used by JLA) for each supernova are plotted. It shows both models almost give the same values of $c$, and the slope of the regression equation by least square method is 1.109.  In Fig.\ref{fig:Dm15_ANN_VS_JLA}, the parameter $\Delta m_{15}$ from the ANNSLCT model and the stretch $X_1$ used in JLA  for each supernova are compared, since they almost play the same role to describe a supernova. From Fig.\ref{fig:Dm15_ANN_VS_JLA}, one can see there is a negative correlation between $\Delta m_{15}$ and $X_1$  due to the definition. Note that the relationship between $\Delta m_{15}$ and $X_1$ is not fully linear. One can also find that our samples have smaller uncertainties than those in the JLA.
	
\section{Cosmological Model Constraint} \label{sec:fit_cosmos}
To constrain the parameters of $\Lambda$CDM, we  minimize the following $\chi^2$ function:
\begin{equation}
	\chi^2 = \chi^2_{SN} + \chi^2_{CMB} + \chi^2_{BAO}\,,
\end{equation}\label{eq:chi2}
by using the data of SN Ia retained from the ANNSLCT model, the Cosmic Microwave Background (CMB) and the Baryon Acoustic Oscillations (BAO) data samples. The $\chi^2_{SN}$ is from the supernova data and it is calculated as
\begin{equation}
	\chi^2_{SN} = (\hat{\bm{\mu}} - \bm{\mu}_{\Lambda CDM}(z;\Omega_m))^\dagger \bm{C}^{-1}(\hat{\bm{\mu}} - \bm{\mu}_{\Lambda CDM}(z;\Omega_m))\,,
\end{equation}
where \bm{$\hat{\mu}$} is calculated by Eq.(\ref{eq:mu}), and \bm{$C$} is covariance matrix of \bm{ $\hat{\mu}$}. We take the same values of the observed peak magnitude $m_B$ and the host stellar mass $M_{stellar}$ as those in the JLA sample. The color $c$ and $\Delta m_{15}$ parameters for each individual supernova are constrained from the template model, and the coefficients $\alpha$ and $\beta$ will be fitted along with the parameters in the $\Lambda$CDM model.

The CMB data that can be measured precisely include $\bm{\nu}_{CMB} = (\Omega_b h^2,\Omega_c h^2, 100\Theta_\ast) ^T$, see Refs.\cite{Bennett:2012zja,Feng:2012gf,Feng:2012gr}.
Here $\Omega_b h^2$ is the baryon density, $\Omega_c h^2 $ is the dark matter density, and $\Theta_\ast$ is the approximation of the sound horizon angular size\cite{Hu:1995en}. The $\chi^2_{CMB}$ of the CMB data is given by:
\begin{equation}
	\chi^2_{CMB} = (\bm{\nu}_{obs} - \bm{\nu}_{\Lambda CDM})^\dagger \bm{C}^{-1}_{CMB} (\bm{\nu}_{obs} - \bm{\nu}_{\Lambda CDM})\,,
\end{equation}
where $\bm{C}_{CMB}$ is covariance matrix.
\par
The BAO data used in Refs.\cite{Feng:2012gr}\cite{Anderson:2012sa} are $\bm{d}_{BAO}=(d_{0.106},d_{0.35},d_{0.57})^T$, and the $\chi^2_{BAO}$ is given by:
\begin{equation}
	\chi^2_{BAO} = (\bm{d}_{obs} - \bm{d}_{\Lambda CDM})^T \bm{C}^{-1}_{BAO} (\bm{d}_{obs} - \bm{d}_{\Lambda CDM})\,,
\end{equation}
where $\bm{C}_{BAO}$ is covariance matrix.

\begin{figure}[t]
	\begin{center}
		\includegraphics[height=0.33\textheight,width=0.5\textwidth,angle=0]{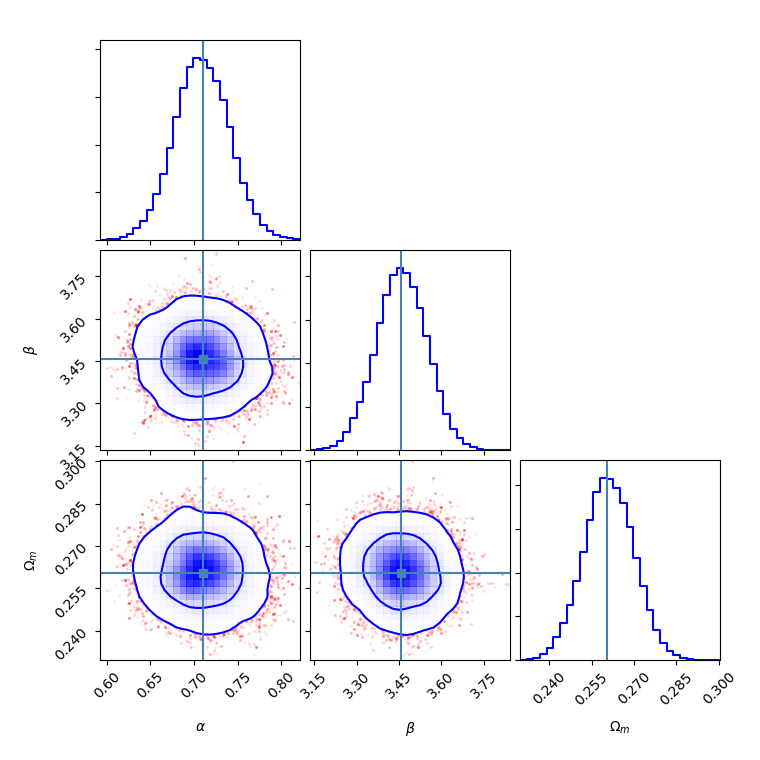}
		\caption{68\% and 95\% confidence contours for the parameters of the $\Lambda$CDM model.}
		\label{fig:contour}
	\end{center}
\end{figure}

The parameters to be constrained include the supernova nuisance parameters $\alpha \text{ or }\alpha', \beta$ , $M_B$ and $\Delta_M$ and the cosmological model parameters $H_0$ and $\Omega_m$. The best-fit values  and $1\sigma$ uncertainties for these parameters are listed in Table.\ref{tab:paras_tabel}. One can see that the two samples give very close values of $\Omega_m$, $H_0$ and other nuisance parameters for the supernova except $\alpha$ and $\alpha'$ because of their different definitions. The contours for $\alpha$, $\beta$ and $\Omega_m$ are plotted in Fig.\ref{fig:contour}.

\begin{figure}
	\begin{center}
		\includegraphics[height=0.28\textheight,width=0.5\textwidth,angle=0]{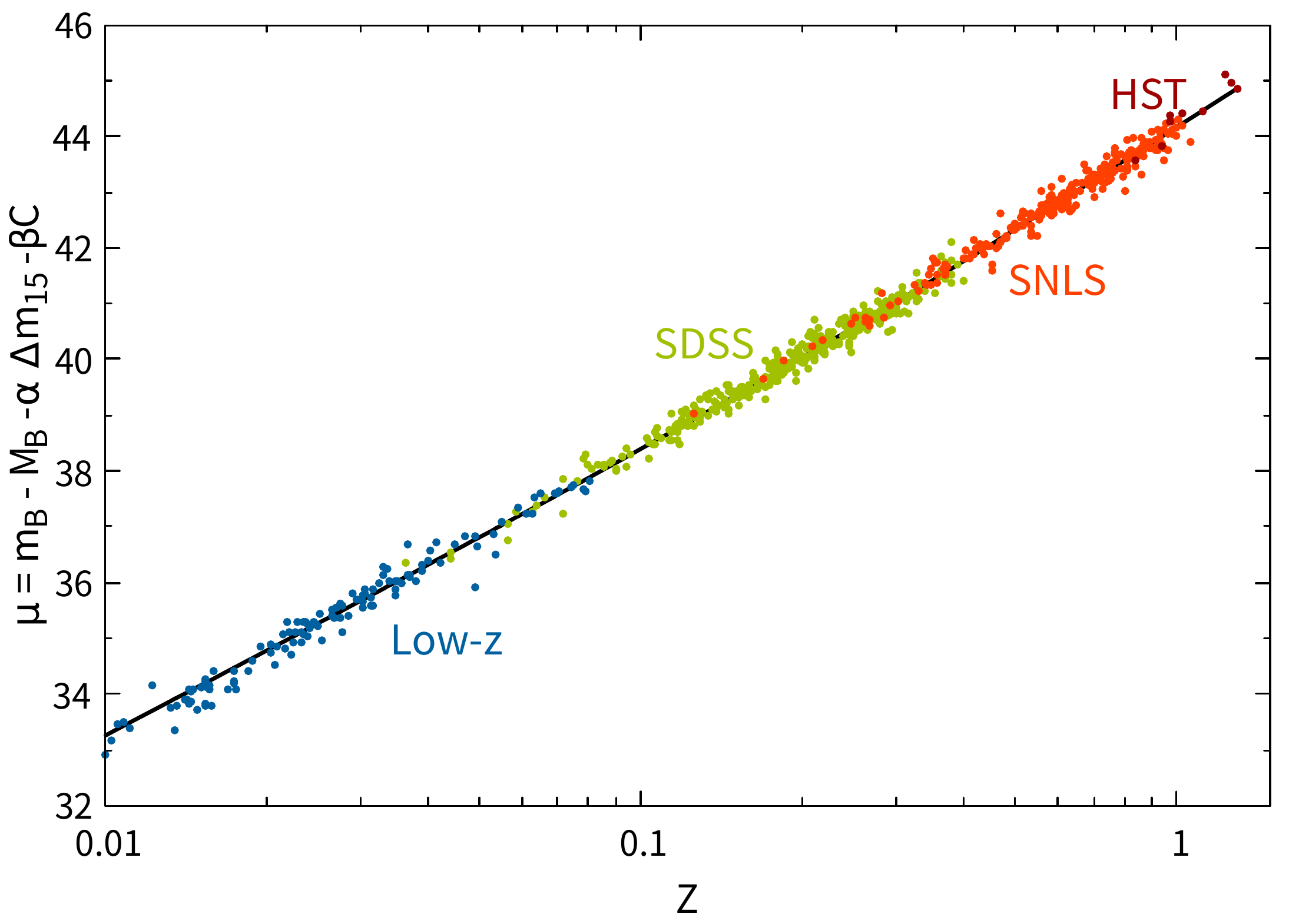}
		\caption{Hubble diagram of the ANNSLCT model. Residuals have been shown by Fig.\ref{fig:residual}.}
		\label{fig:hubble_diagram}
	\end{center}
\end{figure}

The Hubble diagram for the ANNSLCT model is shown in Fig.\ref{fig:hubble_diagram}.
The best-fit values of parameters for the $\Lambda$CDM model are $\Omega_m = 0.261\pm 0.009 $ and $H_0 =67.73\pm 0.70 km s^{-1} Mpc^{-1} $.

\begin{figure}
	\begin{center}
		\includegraphics[height=0.18\textheight,width=0.5\textwidth,angle=0]{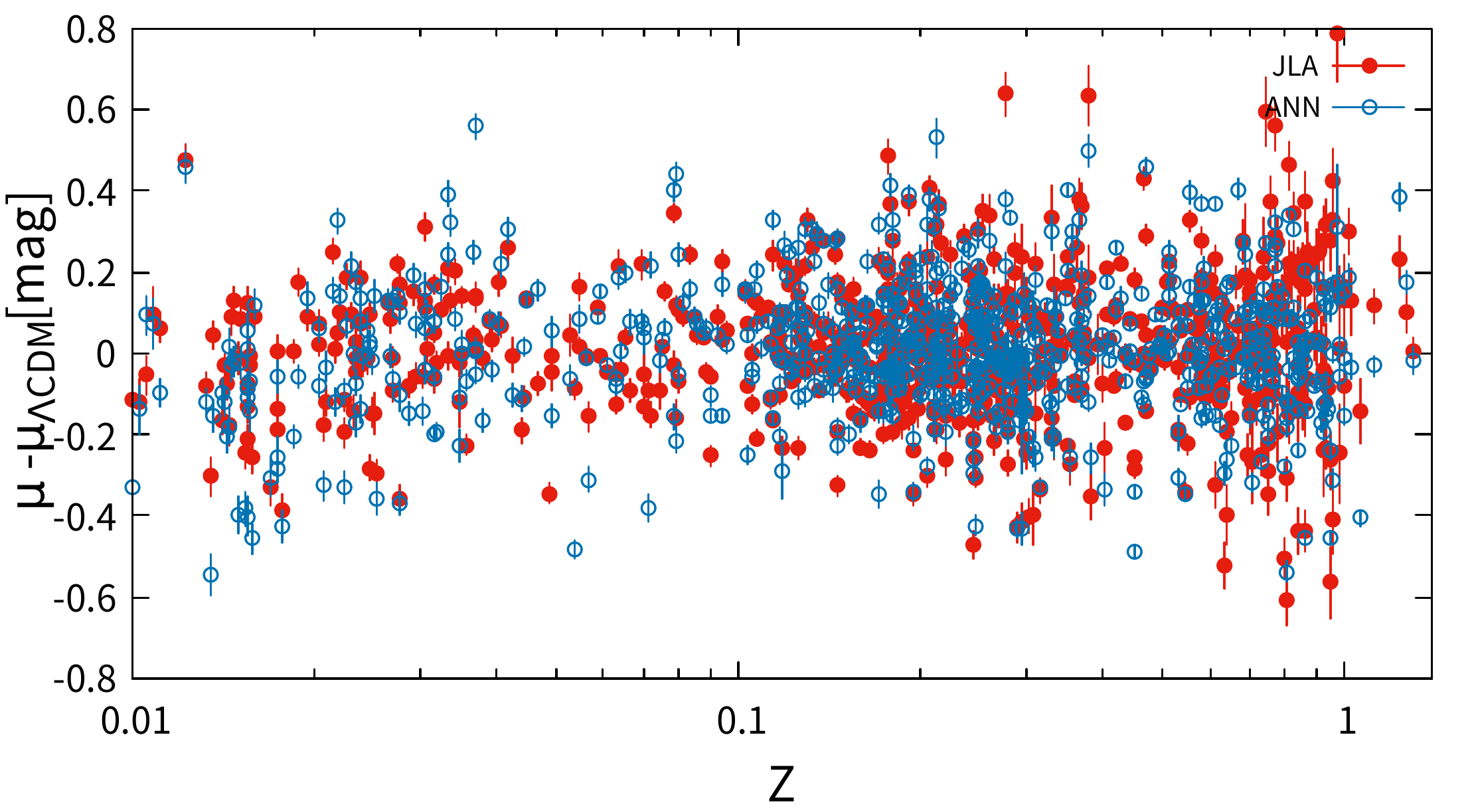}
		\caption{The comparison of residuals with results of the JLA sample retrained by the ANNSLCT  and  SALT2 models.}
		\label{fig:residual}
	\end{center}
\end{figure}

The residuals of two models are shown in Fig.\ref{fig:residual}. 
Specifically, the average absolute values of residual errors are 0.124  mag for the ANNSLCT model and 0.127 mag for the SALT2 model. The distance modulus has a larger bias in the high-z region than that in the low-z. This is partially because the high-z data have low signal-to-noise ratio and poor samples. 

\begin{widetext}
\begin{center}
\begin{table}[!hbtp]
	\begin{tabular}{ccccccccc}
		\hline
		\hline
		Samples & $\Omega_m$ &$H_0$  & $\alpha$ & $\alpha'$ & $\beta$ & $M_B^\ast$  & $\Delta_M$ & $\chi^2/d.o.f$  \\
		\hline
		ANN & 0.261$\pm$0.009 & 67.79$\pm$0.70 &0.710$\pm$0.029 &-& 3.45$\pm$0.083 & -19.18$\pm$0.021 & -0.081$\pm$0.019 & 1062/739 \\
		JLA & 0.257$\pm$0.009 & 67.98$\pm$0.74  &
		-& 0.141$\pm$0.006 & 3.10$\pm$0.074 & -19.11$\pm$0.026 & -0.070$\pm$0.023 & 684/739 \\
		\hline
		\hline
	\end{tabular}
	\caption{Constraints of the  $\Lambda $CDM parameters by two samples.}\label{tab:paras_tabel}
\end{table}
\end{center}
\end{widetext}



\section{Conclusions and Discussions} \label{sec:conclusion}

By using the ANN, we construct the SED sequence for SN Ia. The SED sequence could give both the spectrum with  wavelength range from 3000\AA~to 8000\AA~ and the light curve with phase from 20 days before to 50 days after the  maximum luminosity for the supernovae with different colors and decline rates. In Ref.\cite{Cheng:2018nhz} , the authors have constructed the mean SED sequence with ANN by using the SN Ia spectrum data with and without the color parameter. In this paper, the shape parameter, which is the decline rate $\Delta m_{15}$, is also taken as another input variable for the ANN. The light curves generated by this ANNSLCT model are well consistent with observational data. We find that the model becomes more accurate than that in Ref.\cite{Cheng:2018nhz} when adding the high-z spectra  and light curve data.

Usually, before training the SED sequence model, one needs to assume a functional form for the flux with  color and shape parameters like Eq.(\ref{equ:salt2}). However, there is no need to assume any relations in the ANNSLCT model, since all relations are encoded in the ANN itself. After training, one can obtain not only the SED sequence, but also the parameter values of each supernova, instead of fitting each supernova，as SALT2 did. Another advantage is that the model obtained by using the ANN is automatically differentiable, then it could be easily used to analyze some subsequent physical process. For example, in Ref.\cite{Cheng:2018nhz},  the authors obtained the relation between color and Si II $\lambda$6355 absorption velocity by taking the derivative of the network directly.

We also use the SN Ia sample generated by the ANNSLCT model, CMB and BAO data to constrain the parameters of $\Lambda$CDM model. The best-fit values of parameters for the $\Lambda$CDM model are $\Omega_m = 0.261\pm 0.009 $ and $H_0 =67.73\pm 0.70 km s^{-1} Mpc^{-1} $, which are almost the same as those from the JLA+CMB+BAO joint constraint with the SALT2 model. The average absolute value of residual errors is 0.124  mag for the ANNSLCT model, which is slightly smaller than 0.127 mag for the SALT2 model.

Furthermore, the SED sequence obtained by the ANN model is actually not a linear approximation of the SALT model, so it is expected to improve the SALT2 model. And it may also help us to understand which are the main factors to describe the type Ia supernova explosion, so the ANNSLCT model is worth further study.



\acknowledgments

This work is supported by National Science Foundation of China grant Nos.~11105091,~10671128 and~11047138, the Key Project of Chinese Ministry of Education grant No.~211059,``Chen Guang'' project supported by Shanghai Municipal Education Commission and Shanghai Education Development Foundation Grant No. 12CG51, and Shanghai Natural Science Foundation, China grant No.~10ZR1422000. 

\end{document}